\documentclass[aps,prb,reprint,superscriptaddress, longbibliography]{revtex4-1}
\usepackage{H1}
\usepackage{amsmath}
\usepackage[pdftex,colorlinks=true]{hyperref}
\hypersetup{
citecolor = blue
}
\usepackage[normalem]{ulem}
\usepackage{amssymb}
\usepackage{amsthm}
\usepackage{amsfonts}
\usepackage{bbm}
\usepackage{array}

\begin{document}
\title{Critical Properties of the Many-Body Localization Transition}

\author{Vedika Khemani}
\affiliation{\mbox{Department of Physics, Princeton University, Princeton, NJ 08544, USA}}
\author{S. P. Lim}
\affiliation{\mbox{Department of Physics and Astronomy, California State University, Northridge, CA 91330, USA}}
\author{D. N. Sheng}
\affiliation{\mbox{Department of Physics and Astronomy, California State University, Northridge, CA 91330, USA}}
\author{David A. Huse}
\affiliation{\mbox{Department of Physics, Princeton University, Princeton, NJ 08544, USA}}
\affiliation{\mbox{Institute for Advanced Study, Princeton, NJ 08540, USA}}

\begin{abstract}
The transition from a many-body localized phase to a thermalizing one is a dynamical quantum phase transition which lies outside the framework of equilibrium statistical mechanics.  We provide a detailed study of the critical properties of this transition at finite sizes in one dimension. We find that the entanglement entropy of small subsystems looks strongly subthermal in the quantum critical regime, which indicates that it varies discontinuously across the transition as the system-size is taken to infinity, even though many other aspects of the transition look continuous. We also study the variance of the half-chain entanglement entropy which shows a peak near the transition, and find substantial variation in the entropy across eigenstates of the \emph{same} sample. Further, the 
sample-to-sample variations in this quantity are strongly growing, and larger than the intra-sample variations. We posit that these results are consistent with a picture in which the transition to the thermal phase is driven by an eigenstate-dependent sparse resonant ``backbone'' of long-range entanglement, which just barely gains enough strength to thermalize the system on the thermal side of the transition as the system size is taken to infinity. This discontinuity in a global quantity --- the presence of a fully functional bath --- in turn implies a discontinuity even for local properties. We discuss how this picture compares with existing renormalization group treatments of the transition.
\end{abstract}

\maketitle

\section{Introduction}
Understanding the nature of quantum phases and phase transitions is part of the bedrock of condensed matter physics. The traditional understanding in this field uses the framework of equilibrium statistical mechanics to classify phases according to local patterns of symmetry breaking a la Landau or, more recently, according to various classes of topological order. The transitions between phases---signalled by singularities in thermodynamic functions or observables---are either first-order or continuous, where the latter generally involve a diverging length scale and universal critical scaling behavior.

Progress on the phenomenon of many-body localization (MBL) has revealed the incompleteness of the above framework.  MBL generalizes the
phenomenon of Anderson localization in non-interacting disordered systems to the interacting setting \cite{Anderson58, Basko06, PalHuse, OganesyanHuse, Nandkishore14, AltmanVosk}.  
The transition between many-body localized and thermalizing phases  is not a thermodynamic phase transition, so it need not conform to the usual classifications of phase transitions.  Instead, it is a \emph{dynamical} phase transition between a thermalizing phase which obeys equilibrium thermodynamics in its long time behavior and the MBL phase where the system's dynamics does not bring it to thermal equilibrium. It is also an {\it eigenstate phase transition} \cite{Huse13, PekkerHilbertGlass, Bauer13,Vosk14, Kjall14, Chandran14, Bahri15, Khemani15} across which the nature of the system's (highly excited) many-body eigenstates changes in a singular way from thermal and ``volume-law'' entangled eigenstates that obey the eigenstate thermalization hypothesis (ETH) \cite{Deutsch, Srednicki, Rigol} to non-thermal and area-law entangled eigenstates in the MBL phase.

\begin{figure}[]
  \includegraphics[width=.85\columnwidth]{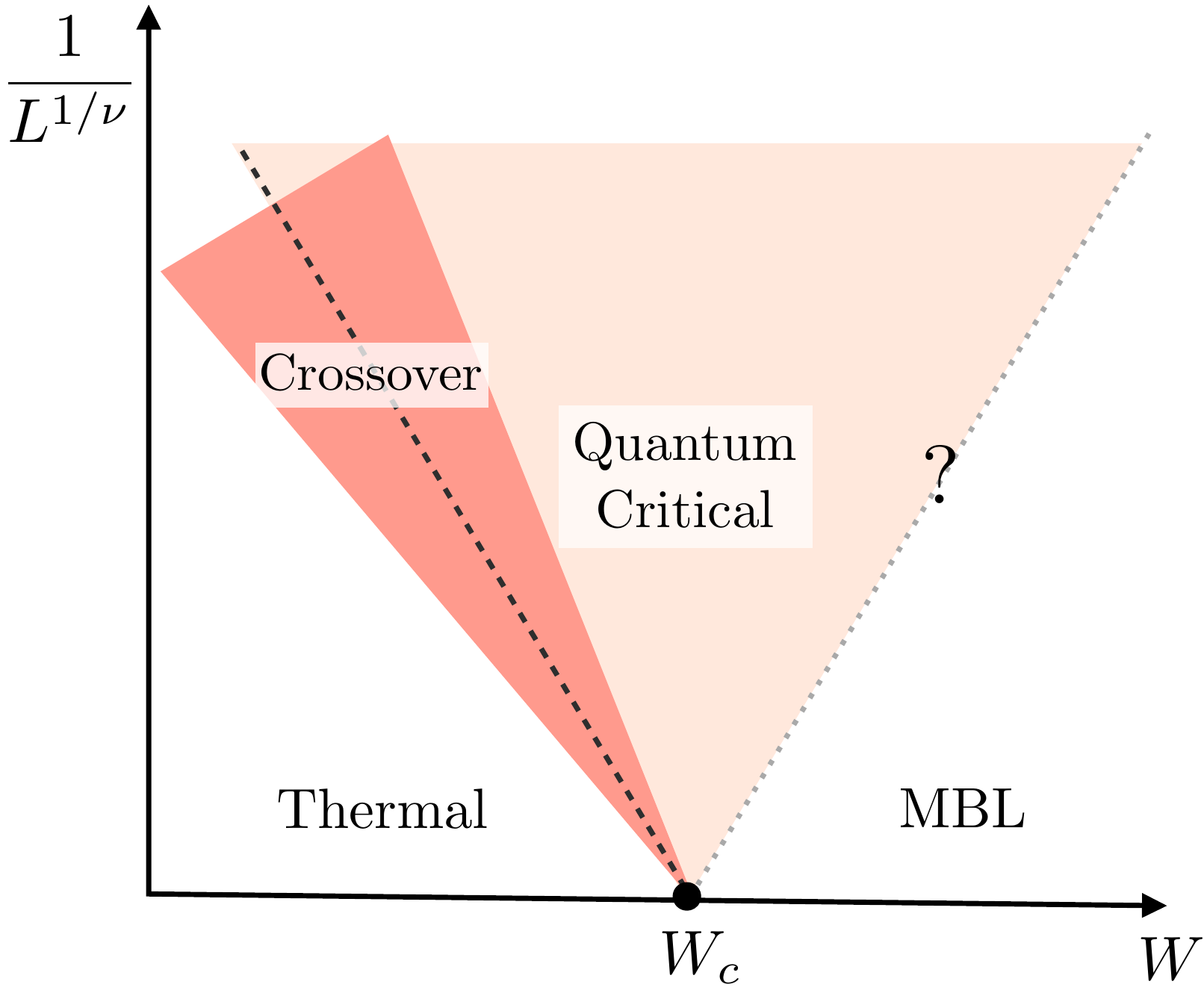}
  \caption{\label{fig:CriticalSchematic} Schematic depiction of the MBL-to-thermal phase transition as a function of disorder strength $W$ and system size $L$, showing the quantum critical regime at finite sizes. Exact diagonalization studies have only observed a crossover on the thermal side of the transition, with no observed crossover between the MBL and quantum critical regimes. }
 \end{figure}

Since the MBL-to-ETH transition lies outside the purview of equilibrium statistical mechanics, very little is definitively known about its properties. 
A recent paper (CLO\cite{CLO}) derived a generalized Harris/Chayes bound\cite{Harris, CCFS1, CCFS2}, $\nu_{FS} \geq 2/d$, for an appropriately defined finite-size correlation length exponent $\nu_{FS}$ associated with the disorder-driven MBL transition in $d$ dimensions.  Recent phenomenological renormalization group (RG) studies in one-dimensional systems find a continuous transition with a localization length exponent $\nu > 2$ satisfying this CLO inequality \cite{VHA,PVP,ZhangRG}. On the other hand, all exact diagonalization (ED) numerical studies to date\cite{PalHuse, Kjall14, Luitz15} (which are limited to small system sizes $L\sim 22$) have found apparent scaling exponents $\nu$ which violate this CLO bound.  Interestingly, all these ED studies have observed a finite-size crossover only on the thermal side of the transition (c.f Fig.~\ref{fig:CriticalSchematic}) with no observed crossover between the MBL and quantum critical regimes. 

In a separate development, Grover formulated an important constraint on the scaling of the entanglement entropy (EE) at the MBL-to-ETH transition\cite{GroverCP}.  Grover considered the entanglement entropy $S_A$ of a small subsystem of length $L_A$ in a much larger system near the phase transition so that $L_A \ll L, \xi$,  where $L$ is the system size and $\xi$ is a diverging correlation length. He made two crucial assumptions: (i)   $S_A$ is a scaling function only of $L_A/\xi$ with no significant $L$ dependence and (ii) $S_A$ varies continuously across the phase transition even after the limit $L\rightarrow\infty$ is taken.  For a conventional continuous transition, these assumptions seem reasonable. Since $S_A$ seems like a {\it local} property of a small subsystem of size $L_A \ll L$, we might not expect it to strongly depend on $L$.  Nor might we expect such a local quantity to be discontinuous across a continuous phase transition. From these assumptions and the strong subadditivity of entanglement, it follows \cite{GroverCP} that $S_A$ must show {\it thermal} volume-law entanglement at the MBL phase transition.

On the other hand, numerical studies of the MBL transition hint at sub-thermal entanglement entropy near the
transition \cite{Kjall14, Luitz15, Devakul15,LimSheng}, although these studies have not focused on this question or the relevant limit $L_A \ll L$.

In the present work, we provide a detailed study of the the MBL-to-thermal phase transition in one dimension, both in the finite-size quantum critical (QC) regime and in the critical-to-thermal crossover regime shown in Fig~\ref{fig:CriticalSchematic}. We show that, contrary to Grover's results, the EE for small subsystems $S_A$ (we use a very small subsystem: one spin) is strongly subthermal in the QC regime---thereby indicating that $S_A$ varies discontinuously across the MBL transition in the limit $L\rightarrow \infty$, a striking result given that many other features of this transition look continuous. 

We also add to the understanding of finite-size scaling at the transition by numerically studying the variance of the half-chain entanglement entropy (EE) which peaks at the MBL-to-thermal transition as the nature of the eigenstates changes from area law to volume law entangled\cite{Kjall14}. We parse in detail the contributions to this variance which come from sample-to-sample, eigenstate-to-eigenstate and cut-to-cut variations. Strikingly, we find a volume law scaling (\emph{i.e.}, a substantial variation) for the standard deviation of the half-chain EE across eigenstates in the \emph{same} sample,  a property which has heretofore not been discussed by any numerical or phenomenological RG treatments of the transition. Further, while the cut-to-cut variations are sub-dominant (and sub volume law), we find that the sample-to-sample variations give the largest contribution to the standard deviation and grow super-linearly with $L$ at the system sizes probed. As we will discuss, this detailed parsing helps us identify the likely source of the observed violations of the CLO inequality and helps us formulate a possible picture of the universal critical properties of the transition. 

Inspired by these data, we present a picture for the finite-size behavior near the phase transition which is consistent with both the discontinuity in $S_A$ and the observed trends in the variance of the half-chain entropy: Essentially, the transition to the thermal phase appears to be driven by a sparse resonant ``backbone'' of long-range entanglement\cite{PVP} which just barely gains enough strength to become a functional ``bath'' and thermalize the entire system in the $L\rightarrow\infty$ limit on the thermal side of the transition.  This corresponds to a discontinuity in a {\it global} quantity---the presence of a fully functional and infinite bath---across the transition. Such a global discontinuity has been observed in other conventional continuous phase transitions, the superfluid density at the Kosterlitz-Thouless transition being an example; the surprising consequence is that, for the MBL transition, this global discontinuity also implies a discontinuity in seemingly local properties like $S_A$.  

Our picture of the transition helps us better understand the nature of the many-body resonances driving the transition, and suggests that the strong-randomness RG analyses in Refs.~\onlinecite{VHA} (VHA) and \onlinecite{ZhangRG} might have made too strong an assumption in allowing only for locally thermalizing and insulating regions, while not permitting something intermediate which is entangled over large distances but is not itself well thermalized.  The RG in Ref.~\onlinecite{PVP} (PVP), on the other hand, is closer to the picture we propose: they allow for sparse resonant clusters of spins in the QC regime that might not fully thermalize the insulating regions spatially interspersed between the resonant spins.  However, in comparing our data to the RG results, we need to keep in mind that the range of sizes we explore numerically are much smaller than the asymptotic regime treated by these RGs.  Thus our picture may apply to an intermediate regime before the asymptotic large-$L$ scaling regime. Nevertheless, it is interesting to note that a more careful reading of the numerical results from PVP's asymptotic RG study actually supports our claim for subthermal $S_A$ (as we will discuss), although PVP do not address or resolve the apparent discrepancy between their data and Grover's constraint. 

We note that a recent preprint \cite{ClarkBimodal} studies the coefficient of the volume law for the EE of subsystems with size $L_A \sim L/4$ and has results both consistent with and complementary to our work. Ref. \onlinecite{ClarkBimodal} finds probability distributions of the entanglement which look increasingly bimodal at the transition; we comment on how their results together with our observed discontinuities suggest that the MBL-to-ETH transition may be some sort of hybrid between continuous and discontinuous phase transitions.

In the balance of the paper, we introduce and benchmark the model used in our analysis (Section~\ref{sec:model}). We then present our numerical data for $S_A$ in Section~\ref{sec:S1plat} and show that it looks strongly subthermal in the quantum critical region. This is followed by a finite-size scaling analysis for $S_A$ in \ref{sec:S1scaling}, together with a comparison to Grover's results. In Section~\ref{sec:variance}, we study the variance of the half chain EE and parse the contributions coming from fluctuations across samples, eigenstates and spatial cuts. In Section~\ref{sec:picture} we sketch a picture of the transition consistent with our observations, and end with a summary and outlook in Section~\ref{sec:conclusion}.

\section{The Model}
\label{sec:model}
We study a spin-1/2 Heisenberg chain with random $z$-fields and nearest and next-nearest neighbor interactions:
\begin{align}
H &= J\sum_{i=1}^{L-1} [(S_i^xS_{i+1}^x + S_i^y S_{i+1}^y) + S_i^z S_{i+1}^z] + \sum_{i=1}^{L} h_i S_i^z \nonumber\\
&+  J'\sum_{i=1}^{L-2} (S_i^xS_{i+2}^x +S_i^yS_{i+2}^y ) ~,
\label{eq:model}
\end{align}
where $S_i^{\{x/y/z\}}$ are spin 1/2 degrees of freedom on site $i$, $J=J'=1$ and the fields $h_i$ are drawn uniformly and independently from  $[-W,W]$. This model is MBL for large disorder strength 
$W>W_c \geq 7$.  We present the estimate of $W_c$ as a lower bound since, as usual, 
we do not observe a 
crossover on the MBL side of the transition. 

Note that this model with $J'=0$ is a ``canonical'' model used in many MBL studies with a critical $W_c \geq 3.5$.\cite{Luitz15, Devakul15}  We found it prudent to add the next-nearest neighbor term to break the integrability of the canonical model in the limit $W \rightarrow 0$.  Since our goal is to discriminate between thermal and sub-thermal scaling for the critical EE, it helps to have the MBL phase abut a strongly-thermalizing phase.  In the canonical model, the EE does not reach the thermal value until relatively deep in the delocalized phase (for numerically accessible system sizes), thus making it problematic to draw meaningful conclusions about an observed subthermal critical EE. Due to not being integrable at $W=0$, our model thermalizes more completely within the thermal phase for the smallest system sizes in our study.

\begin{figure}[]
  \includegraphics[width=\columnwidth]{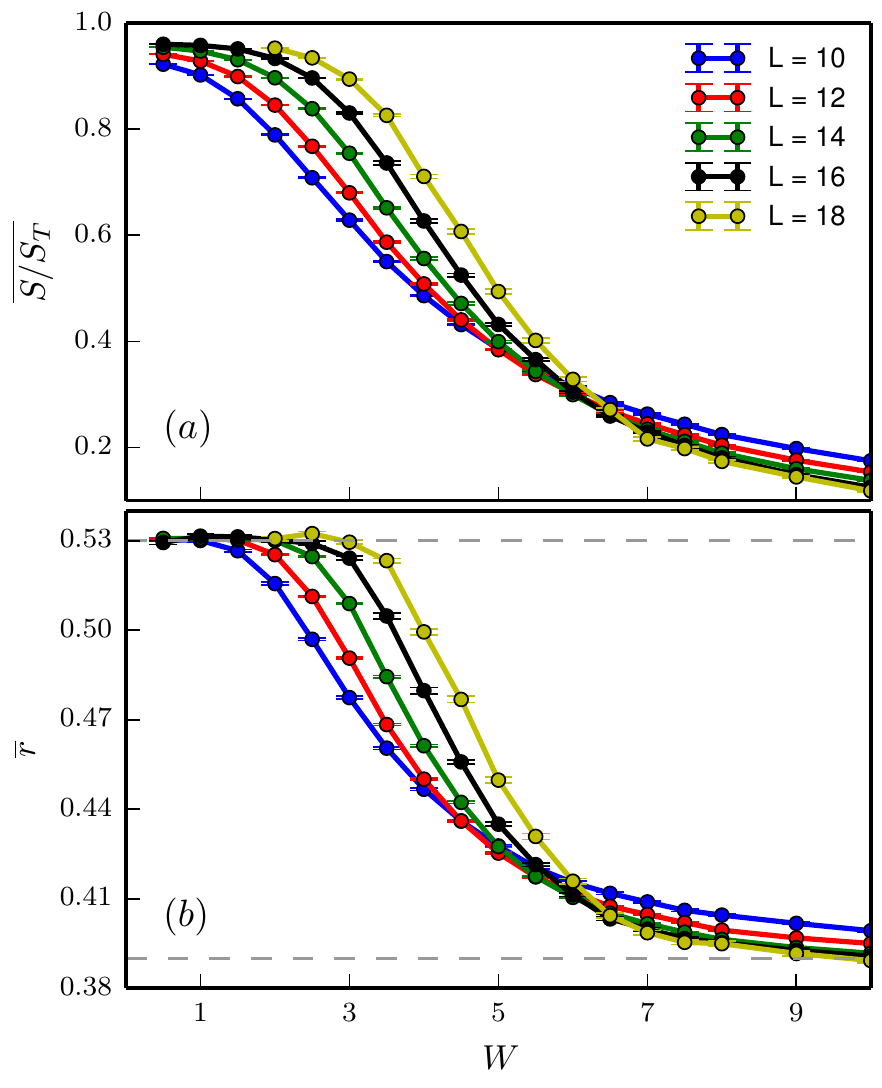}
  \caption{\label{fig:NNN}  (a) Disorder averaged half-chain entanglement entropy divided by the Page value $S_T$ for a random pure state as a function of $W$ and $L$. $S/S_T$ approaches a step function at the transition, going from zero in the MBL phase with area-law entanglement to one in the thermal phase. (b) Disorder averaged level statistics ratio $\bar r$ which obeys GOE/Poisson distributions in the thermal/localized phases, respectively.}
 \end{figure}

Fig.~\ref{fig:NNN} benchmarks the location of the transition in \eqref{eq:model} using the half-chain entanglement entropy, $S$, and the level statistics ratio, $r$. Fig.~\ref{fig:NNN}(a) shows $S$ divided by 
$S_T = 0.5(L-\log_2{e})$ bits which is the Page\cite{Page} value for a random pure state. The data is averaged over $2000-10^5$ disorder realizations depending on $L$. Within each sample, the data is further averaged over the 100 eigenstates closest to the center of the band in the $S^z_{\rm tot} = 0$ sector (or a quarter of that sector's Hilbert space for small system sizes). Unless otherwise mentioned, these parameters apply to all our numerical results.  $S/S_T$ as a function of $W$ approaches a step function with increasing $L$, going from zero in the MBL phase with area-law entanglement to one in the thermal phase.

Fig.~\ref{fig:NNN}(b) shows the level statistics ratio $ r \equiv \min\{\Delta_n, \Delta_{n+1}\}/\max\{\Delta_n, \Delta_{n+1}\}$, where $\Delta_n = E_n - E_{n+1}$ is the spacing between eigenenergy levels. This ratio is a sensitive test of the level repulsion in a system: it approaches the GOE (Gaussian Orthogonal Ensemble) value $r \cong 0.53$ in the thermal phase and the Poisson value $r \cong 0.39$ in the localized phase. Figure~\ref{fig:NNN}(b) shows that the system looks nicely thermal at small $W$ and localized at large $W$, with the location of the crossing drifting towards larger $W$ with increasing $L$ as is typical.

\section{``Local'' Entanglement Entropy}

We now turn to the entanglement entropy $S_A$ of subsystems $A$ with length $L_A$ in the limit $L_A \ll \xi, L$.  Given the limited system sizes accessible to an ED study, we choose $L_A = 1$ to make the subsystem as small as possible when compared to the system size $L$.  Fig.~\ref{fig:S1} shows the disorder and eigenstate averaged entanglement entropy $S_1$ (in bits) computed in eigenstates of the full system with the subsystem consisting of one spin at the end of the chain (Appendix~\ref{app:S1dist} shows distributions of $S_1$ instead of just the mean values).   While the data qualitatively look similar for any single-site subsystem in the chain, we use the end spin because the features we want to emphasize in our discussion are the clearest for the end spin.

\subsection{Subthermal plateaus}
\label{sec:S1plat}

\begin{figure}[]
  \includegraphics[width=\columnwidth]{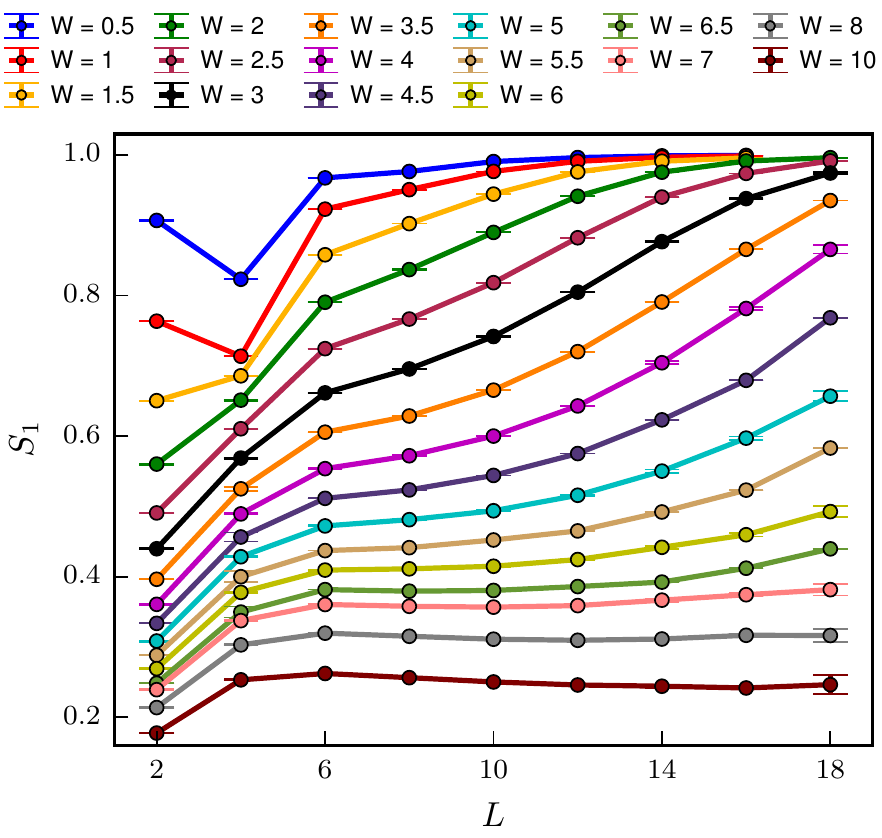}
  \caption{\label{fig:S1} Disorder and eigenstate averaged entanglement entropy $S_1$ (in bits) computed in eigenstates of the full system, for a subsystem comprising one spin at the end of the chain. The (rounded) ``plateaus'' in $S_1$ for intermediate $L$ and $W$ are to be associated with the quantum critical regime and show strongly subthermal values of $S_1.$ }
 \end{figure}

 \begin{figure}[]
  \includegraphics[width=.85\columnwidth]{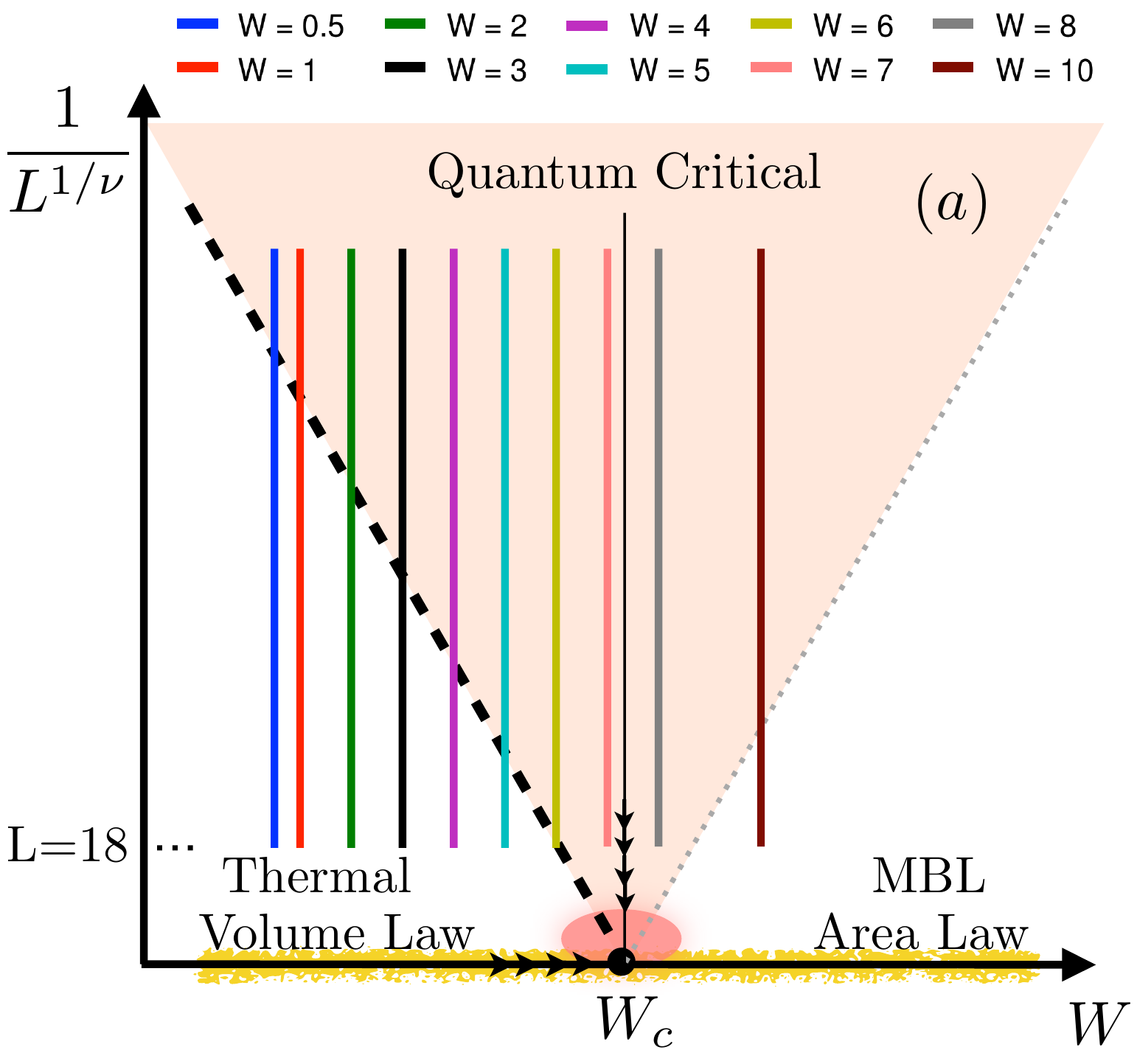}
   \includegraphics[width=.85\columnwidth]{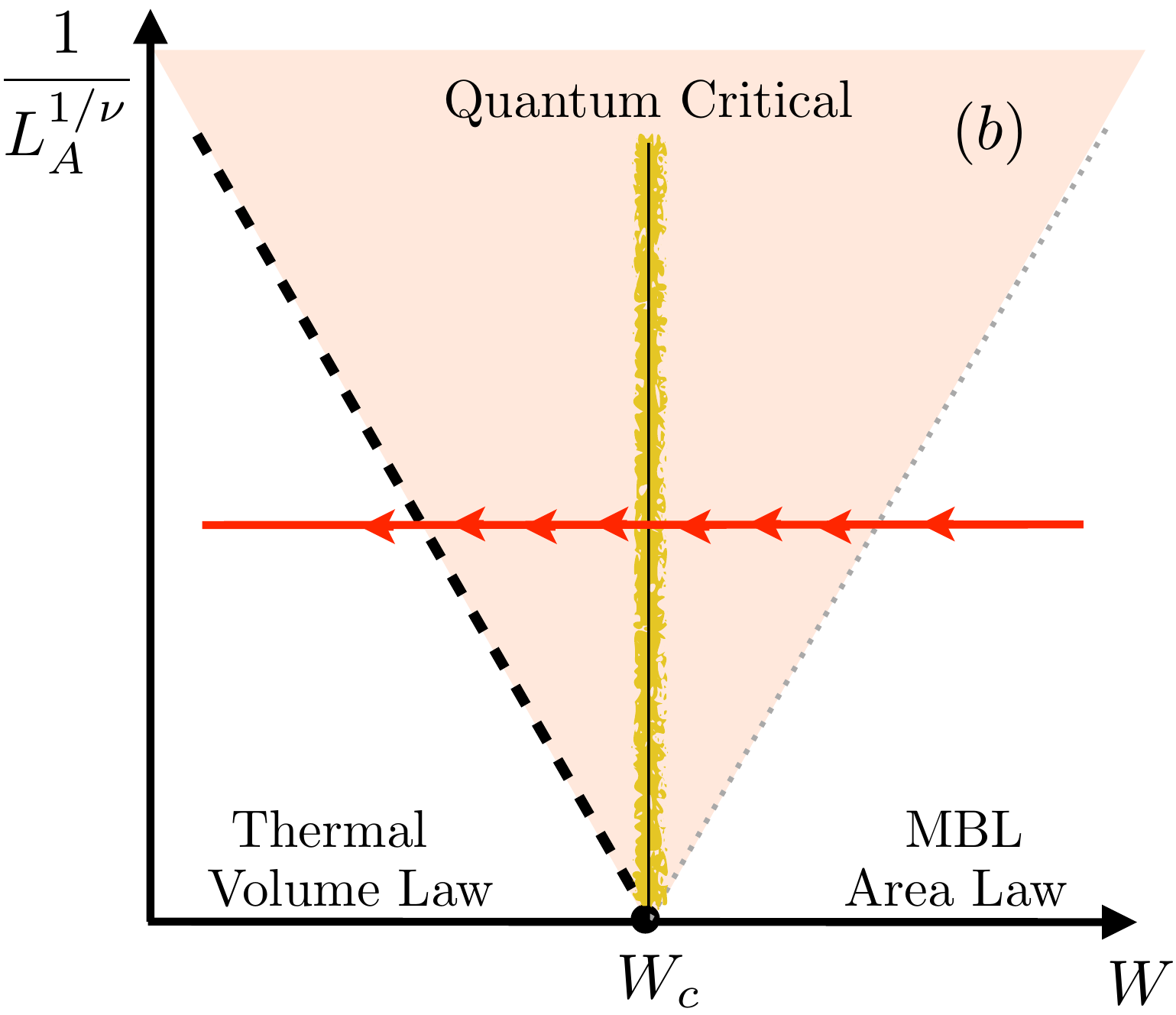}
  \caption{\label{fig:Critical} The MBL-to-thermal phase transition and finite-size crossovers as a function of (a) $L/\xi$ and (b) $L_A/\xi$ (schematic). For the single-site entropy $S_1$, the only relevant scaling variable is $L/\xi$ and the curves at fixed $W$ in Fig. \ref{fig:S1} correspond to the vertical lines in the crossover phase diagram (a) with the subthermal plateaus lying in the QC regime. Grover's analysis\cite{GroverCP} considers $L_A/\xi$ as the relevant scaling variable (b) and shows that if $S_A$ is continuous across $W_c$ then it must be thermal in the critical regime.  The inconsistency between the pictures (a) and (b) can be reconciled if there is a discontinuity in $S_A$ at $W_c$ in the limit $L\rightarrow\infty$ (yellow lines).   }
 \end{figure}

If we assume a continuous transition with some notion of critical finite-size scaling\cite{CLO,GroverCP}, each
value of $W$ defines a correlation length $\xi(W)$ which diverges as the transition at $W_c$ is approached\footnote{We note that the best
definition of this length $\xi$ on either side of the phase transition is still an open question, and there is reason to believe that the
MBL phase is characterized by multiple length scales which diverge differently (or not at all) as the transition is approached\cite{Huse14,KH}.}.
As we increase $L$, if the system follows finite-size scaling, it looks quantum critical for system
lengths $L < \xi$ and should look, respectively, thermal (localized) for $W< W_c$  ($W> W_c$) and $L > \xi$.

Fig.~\ref{fig:S1} shows that deep in the thermal phase (small $W$, small $\xi$), as $L$ is increased $S_1$ quickly approaches one bit of entanglement as appropriate for infinite-temperature thermal eigenstates.  As $W$ is increased towards $W_c$, the correlation length increases and the finite-size effect on $S_1$ gets stronger, since we need to increase the system size to $L \gg \xi$ before $S_1$ approaches its thermal value of one bit.  For a range of $W$ on the thermal side of the phase transition, the evolution of $S_1$ vs. $L$ shows three regimes within the sizes we can access:
\begin{enumerate}
\item At the smallest $L<6$ there is an increase of $S_1$ vs. $L$ that is due to short-range entanglement and is present even within the MBL phase at $W=10$.  We assume that this very small $L$ behavior reflects short range physics and is not in any scaling regime associated with the phase transition.
\item There is an intermediate range of $L$ where an approximate ``plateau'' in $S_1$ vs $L$ starts to develop for $ 4< W < 7$. Note that even though there is no strict ``plateau'' in $S_1$ for the smaller $W$s, the evolution of the curves clearly shows the development of an extended range where $S_1$ grows very slowly with $L$, and the length of this approximate plateau grows as the transition is approached. 
\item At larger $L$ for the same disorder range, $4< W < 7$, we observe a stronger increase in $S_1$ vs $L$. This is likely due to the system approaching full thermalization at even larger (inaccessible) system sizes.
\end{enumerate}
In this interpretation, the ``plateau'' in $S_1(L)$, best illustrated at $W=6.5$, is the quantum critical (QC) regime where we are in the thermal phase, but $L < \xi(W)$. 
We can see that the value of $S_1$ on this plateau is substantially less than one bit and is hence strongly sub-thermal. 
Thus, in the QC regime {$L_A \ll L < \xi$}, the eigenstate entanglement entropy is well below its thermal value, in contrast to the conclusion following from Grover's assumptions \cite{GroverCP}.  We will suggest below a scenario where the transition is in some sense discontinuous and thus violates one of those assumptions. Finally, we associate the stronger increase in $S_1$ at larger $L$ with the crossover from the QC regime to thermal phase. 

\subsection{Finite-size scaling for $S_A$}
\label{sec:S1scaling}

We now develop a possible scaling theory of $S_A$ in the vicinity of the MBL  transition.
A general finite-size scaling form for $S_A$ takes the usual form
\begin{equation}
S_A = {L_A} \; f(L^{1/\nu} \delta, L_A^{1/\nu} \delta)
\label{eq:S1Scaling}
\end{equation}
where $\delta = (W-W_c)$, and the lengths are scaled as usual with the correlation length exponent $\nu$, so, for example, $L^{1/\nu}\delta \sim |L/\xi|^{1/\nu}$.  The prefactor of $L_A$ outside of the scaling function is required to match to the volume-law entanglement in the thermal phase.  We now consider two limits: (a) when $L_A \ll L$, the relevant scaling variable {on the thermal side of the transition} is $L/\xi$ and (b) when $L\rightarrow \infty$, the {remaining} scaling variable is $L_A/\xi$.   Fig.~\ref{fig:Critical} shows the finite-size critical crossover regime as a function of the relevant scaling variables in these two limits. 

For the data in Fig.~\ref{fig:S1}, $L_A = 1$ and thus $L/\xi$ is the relevant scaling variable, as in Fig.~\ref{fig:Critical}(a). The ``plateaus'' of $S_1(L)$ correspond to the critical and MBL regimes, with the crossover between these two regimes undetected as usual (grey dotted line in Fig.~\ref{fig:Critical}(a)).  The gradual crossover between the critical and thermal regimes is the stronger increase of $S_1(L)$ with increasing $L$ from the plateau value towards the thermal value of one bit; we indicate its approximate location with the dashed line in Fig.~\ref{fig:Critical}(a).  Tuning $L$ at fixed $W$ corresponds to taking vertical cuts in the crossover phase diagram.

On the other hand, in Grover's analysis\cite{GroverCP}, the system size $L \gg L_A, \xi$ and $L_A/\xi$ is the relevant scaling variable. In the limit $L\rightarrow\infty$, the phase transition occurs at $W_c$ even for finite $L_A$. It is clear that $S_A$ must obey thermal volume law scaling on the thermal side of the transition, $W < W_c$, for $L_A \ll L$. Grover's analysis tunes through the MBL transition starting from the localized side as shown in the horizontal cut in Fig.~\ref{fig:Critical}(b).  If we assume that $S_A$ remains continuous throughout this scan in the limit $L\rightarrow\infty$, then his analysis shows that the $S_A$ must be thermal in the QC regime $\xi \gg L_A$, even on the MBL side of the transition. For a typical continuous thermodynamic phase transition, a local quantity like $S_A$ is indeed continuous through the transition.  The numerical evidence for subthermal $S_A$ in the QC regime, in contradiction with Grover's conclusion, thus suggests that the assumption of continuity may be incorrect, \emph{i.e.} the transition looks discontinuous if one examines the behavior of $S_A$ through the transition in an infinite system.

In Fig.~\ref{fig:Critical}(b) this discontinuity in $S_A$ will be present for all $L_A$ in the scaling regime along the vertical line at $W=W_c$ (shaded yellow) when $L\rightarrow\infty$. In Fig.~\ref{fig:Critical}(a),  the discontinuity is only present on the horizontal axis (again shaded yellow) where the system size $L$ is infinite. Stated differently, the two limits (black arrows in Fig.~\ref{fig:Critical}(a)) $$\lim_{ W\rightarrow W_c} \lim_{ L\rightarrow \infty}  S_1 \neq \lim_{ L\rightarrow \infty} \lim_{ W\rightarrow W_c  } S_1 $$ may not commute on the thermal side of the transition.  

We should inject a note of caution before concluding this section.  As we argue below, there is evidence that the system sizes accessible to ED studies are not in the asymptotic finite-size scaling regime.
Thus, there remains the possibility that our observation of subthermal $S_A$ for $L_A \ll L$ might be a preasymptotic feature that could change if larger $L$ could be accessed. Thus, it is useful to compare our results with the RG approaches to the transition which study much larger system sizes.
VHA\cite{VHA} do explicitly look at this {question}, but find that their results are too near the boundary between having and not having a discontinuity in $S_A$ to be sure.  PVP\cite{PVP}, on the other hand, invoke Grover's thermal scaling at several points in their paper. However, a more careful reading of their data actually suggests such a discontinuity because they find that only a small fraction of the spins are in entangled resonant clusters at the transition. Thus, a typical subsystem will not lie on the sparse network of thermally entangled clusters and, on average, $S_A$ will look subthermal at the transition. The authors, however, do not address or resolve the apparent discrepancy between their data and Grover's results. It is interesting to note that our data in Section~\ref{sec:variance} lends support to PVP's picture of the transition over VHA's, and thus indirectly further bolsters our claim for subthermal $S_A$. 

\section{Variance of the half-chain entanglement entropy}
\label{sec:variance}
We now switch directions and look at a complementary quantity {that we use, in particular, to examine the \emph{crossover}} between the thermal and quantum critical regimes. The standard deviation of the half-chain entanglement entropy, $\Delta_S$, has been used as a diagnostic for locating the MBL-to-ETH transition\cite{Kjall14,Luitz15}. This quantity shows a peak at the crossover as the eigenstate entanglement changes from thermal to strongly subthermal, while it tends to zero deep in the MBL/ETH phases where the EE for almost all states is area-law/thermal volume law.

\begin{figure}[t]
  \includegraphics[width=\columnwidth]{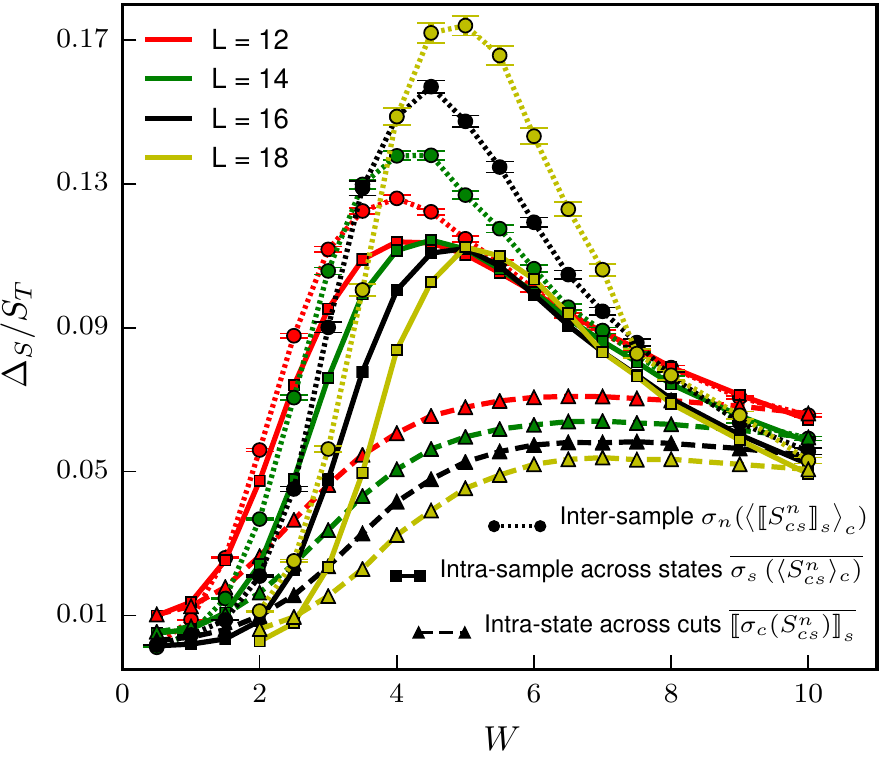}
  \caption{Standard deviation of the half-chain entanglement entropy $\Delta_S$ divided by the random pure state value
$S_T$, parsed by its contributions from cut-to-cut (dashed), eigenstate-to-eigenstate (solid) and sample-to-sample (dotted) variations.}
\label{fig:Variance}
 \end{figure}
 
We parse in detail the contributions to $\Delta_S$ coming from sample-to-sample, eigenstate-to-eigenstate and cut-to-cut variations. Let us denote by $S^{n}_{cs}$ the half-chain EE in a specified eigenstate ``$s$'' in sample ``$n$'' and for a particular bipartite entanglement cut ``$c$'' (which defines the subsystem as extending from some site $i$ to $i+(L/2)-1$). We define $\langle O\rangle_c$, $\llbracket O\rrbracket_s$, and $\overline{O}$ as the average of $O$ with respect to spatial cuts $c$, eigenstates $s$ and disorder samples $n$, respectively. Finally, $\sigma_{\{c/s/n\}}(O)$ represents the standard deviation of  $O$ on varying the $c/s/n$ index.  We use all cuts that fit in the sample length $L$, while we use only the 100 eigenstates closest to zero energy. In Figure~\ref{fig:Variance}, we plot $\Delta_S$ parsed three different ways:
\begin{enumerate}
\item $\Delta_S^{\rm samples}$ = $\sigma_n \left(\langle \llbracket S_{cs}^n\rrbracket_s\rangle_c\right)$ (dotted lines) is obtained by first averaging the half-chain EE over all spatial cuts and eigenstates in a given sample, and then taking the standard-deviation of the averaged entropy across samples. This quantity denotes the sample-to-sample variation in $S_{cs}^n$.
\item $\Delta_S^{\rm states}$ = $\overline{\sigma_s \left(\langle  S_{cs}^n\rangle_c\right)}$ (solid lines) is obtained by taking the standard-deviation of the cut-averaged EE across eigenstates in a given sample, and then averaging over samples. This quantity denotes the eigenstate-to-eigenstate variation in $S_{cs}^n$.
\item $\Delta_S^{\rm cuts}$ = $\overline{\llbracket \sigma_c\left(S_{cs}^n\right)\rrbracket_s}$ (dashed lines) is obtained by taking the standard deviation across spatial cuts $c$ in a given eigenstate of a given sample, and then averaging over eigenstates and samples. This quantity denotes the cut-to-cut variation in $S_{cs}^n$.
\end{enumerate}
We clearly see that, at these sizes, the sample-to-sample variations are larger than the intra-sample variations over eigenstates or cuts.  All three measures of $\Delta_S$  are divided by the thermal entropy $S_T=0.5(L\ln(2)-1)$ bits. Since $S/S_T$ lies between $0$ and $1$, $\Delta_S/S_T$ can be at most 0.5, the value corresponding to a binary distribution of $S$.

 \begin{figure}[t]
  \includegraphics[width=\columnwidth]{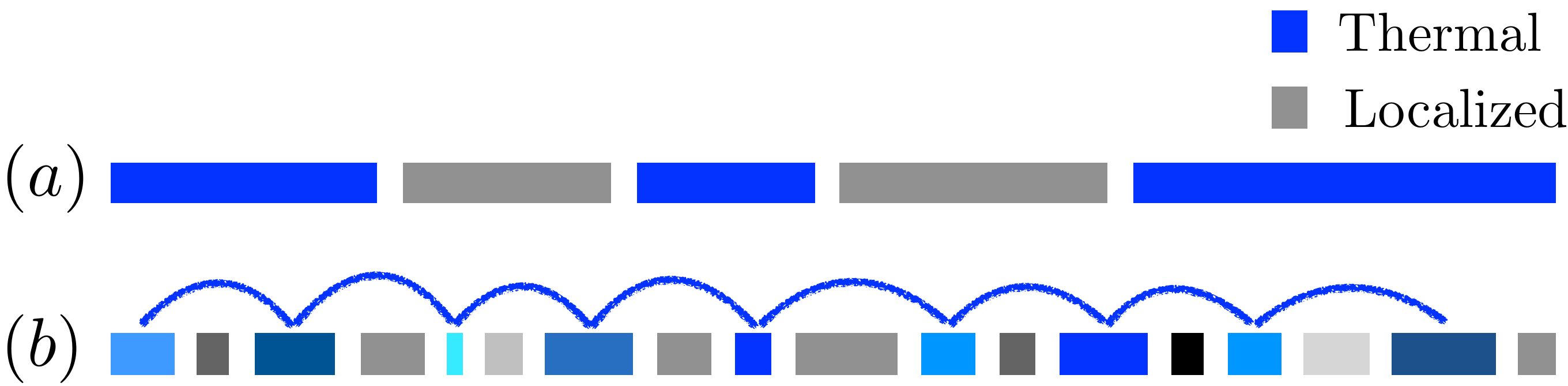}
  \caption{Schematic depiction of two possible models of the crossover from the MBL phase to the thermal phase. (a) The picture from VHA's RG\cite{VHA} predicts contiguous thermal/localized blocks. At the crossover, a few long thermal blocks occupy a finite fraction of the system giving a subthermal but volume-law half-chain EE.  (b) An alternate picture for the transition with a sparse entangled backbone of small thermal blocks of spins with varying degrees of entanglement. The backbone is not contiguous, but spans the entire system and the thin blue lines denote entanglement between the blocks. In both pictures, the thermal clusters acquire just enough strength to thermalize the entire system on the thermal side of the transition in the limit $L\rightarrow \infty$. }
\label{fig:CrossoverModels} 
\end{figure}

First, we note the striking result that the peak value of $\Delta_S^{\rm states}/S_T$ is independent of $L$ indicating a volume law scaling, $\Delta_S^{\rm states} \sim L$, and thus a substantial variance in the half-chain EE across eigenstates in the \emph{same} sample. This property has not been previously noted, nor has it been included by any of the phenomenological RG approaches to the transition. It indicates that the network of resonances driving the transition varies substantially across eigenstates of a given sample, a potentially important feature that deserves further exploration.

Further, the  peak value of $\Delta_S^{\rm samples}/S_T$ grows strongly with $L$ which would naively indicate that  $\Delta_S^{\rm samples} \sim L^{\alpha}$ with $\alpha > 1$. However, since the maximum possible value of $\Delta_S$ is $0.5S_T \sim L$, this super-linear growth is clearly not sustainable in the asymptotic large $L$ limit. This indicates that the observed finite-size violations of Harris/Chayes/CLO bounds (which are derived from sample-to-sample variations) might result from a scenario in which the effect of quenched randomness across samples is not yet fully manifest, but growing strongly, at the sizes studied. Our analysis hints at the possibility of two asymptotic fixed points governing transitions between MBL and thermal phases: one dominated by ``intrinsic'' eigenstate randomness within a given sample, and the second dominated by external randomness which varies across samples. In this framework, the critical scaling collapses in the finite-size systems studied thus far\cite{Kjall14, Luitz15} appear to be in a preasymptotic regime described by the first fixed point (for which Harris/Chayes type bounds do not apply) en route to flowing towards the second.

Finally, note that the peak value of $\Delta_S^{\rm cuts}/S_T$ decreases with increasing $L$, and a scaling analysis (not shown) in fact shows  $\Delta_S^{\rm cuts} \sim L^{1/2}$. 
This scaling sheds light on the potential nature of the many-body resonances driving the transition, and discriminates between the VHA and PVP RG approaches. The VHA\cite{VHA} RG treatment produces a subthermal half-chain EE at the crossover from 1-2 large thermal blocks whose length scales extensively with $L$ (see Fig.~\ref{fig:CrossoverModels}(a) for an illustration).  This picture predicts a cut-to-cut standard-deviation which scales as $\sim L$ at the crossover and is inconsistent with our $\Delta_S^{\rm cuts}$ data at these sizes.  On the other hand, a picture of a sparse network of resonances that is strongly inhomogeneous only on the microscale (Fig.~\ref{fig:CrossoverModels}(b)) is more similar in spirit to the presentation of PVP's RG\cite{PVP} and is consistent with the observed scaling, as we discuss in the next section.

To summarize, we have seen a substantial volume-law scaling for $\Delta_S$ across eigenstates of the same sample. Moreover, the intra-sample variations at these sizes are smaller than the sample-to-sample variations which show strong finite-size effects and unsustainable trends with $L$. We note that Ref.~\onlinecite{ClarkBimodal} also studied the standard deviation across eigenstates, albeit of a different physical quantity in a different model, and they found a small increase in this quantity with $L$. However, even for their data, the sample-to-sample variations increase with $L$ much more strongly than state-to-state variations.

The clear indication from the sample-to-sample data that we are not in the asymptotic large-$L$ scaling regime is perhaps connected to a feature of the critical RG fixed points in VHA and PVP. At these fixed points, the fraction of the sample that is in the entangled clusters is very small: $\sim 1\%$ or less\cite{VHA,PVP}.  If this is an accurate picture of the asymptotic QC regime, then this just can not apply to samples with well under 100 spins since then the entangled clusters would be smaller than one spin.  But the physics on length scales over 100 spins is on time scales over $2^{100}$, so might remain inaccessible to both experimental and numerical work.  Thus the pre-asymptotic QC regime explored by the numerically accessible 
smaller $L$ samples might be closer to what is physically relevant.  We also note that recent papers have applied the density matrix renormalization group technique for studying both the MBL\cite{KhemaniDMRGX, YuPekkerClark, LimSheng} and thermal\cite{Anto2016} phases in disordered spin systems at much larger system sizes than those accessible to ED, although such techniques cannot yet access the MBL-thermal transition. 

One last point. Many disordered statistical physics models look ``self-averaging'' in that the spatial variations between subregions within a large sample are similar to the sample-to-sample variations of smaller samples of the size of those subregions.  This is a type of locality, where the properties of a subregion are not very sensitive to the size of the full sample or to the properties of non-adjacent subregions of the same sample.  But the physics of the MBL transition seems likely not to have this ``locality'':  If some parts of a large sample are such that they locally thermalize and form a good bath, then they may be able to thermalize the entire sample and thus make all subregions strongly entangled.  So the local entanglement properties of a given subregion can be strongly affected by non-adjacent subregions of the sample.

\section{Heuristic Model for the Quantum Critical and Crossover Regions}
 \label{sec:picture}
We now present a picture of the quantum critical and crossover regimes that is consistent with the observed scaling for $S_1$ and $\Delta_S$.
{We also comment on how this picture compares to the VHA and PVP RG frameworks\cite{PVP, VHA}.}

\begin{figure}[t]
  \includegraphics[width=\columnwidth]{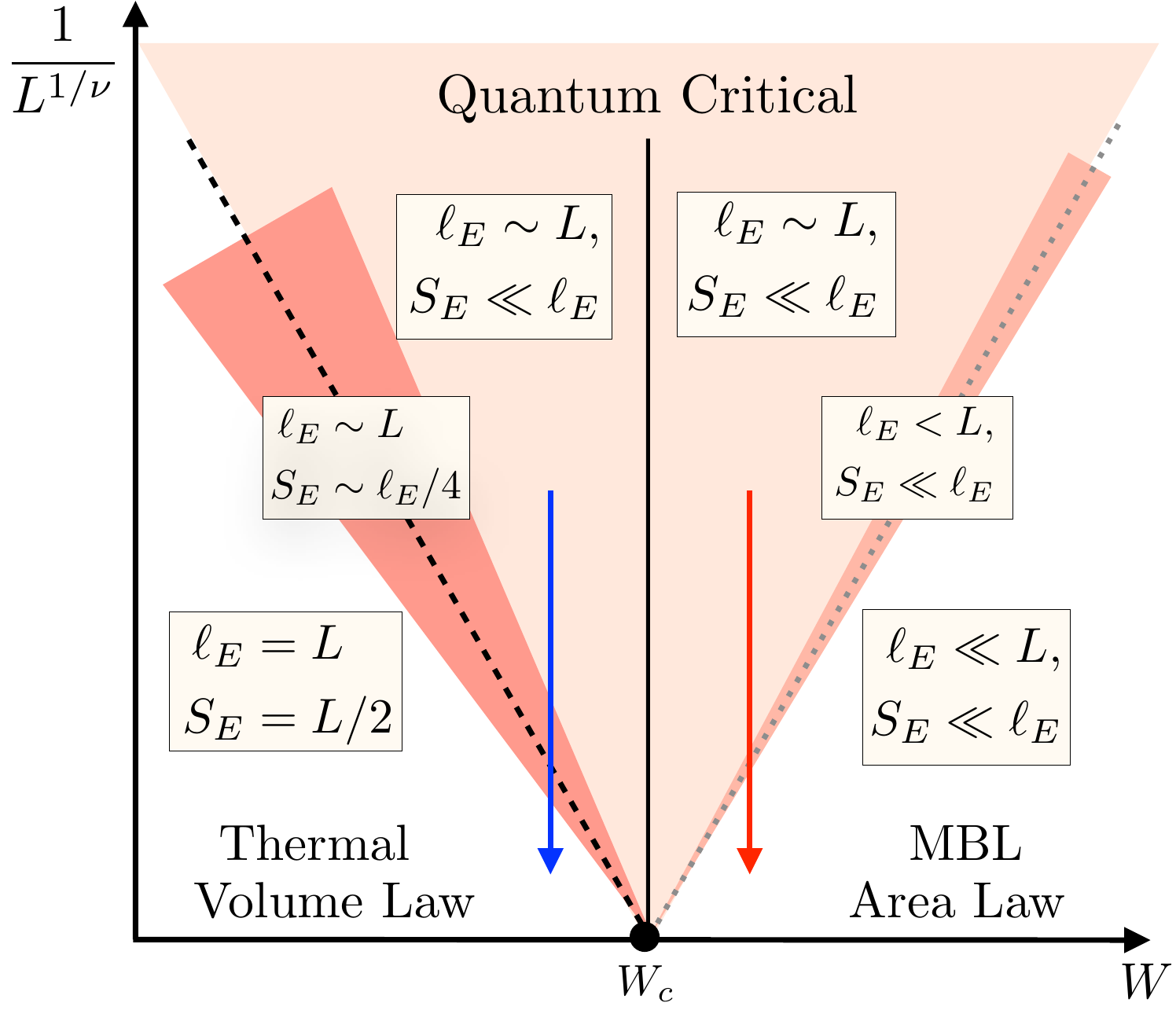}
  \caption{Schematic illustration of our ``picture'' of the MBL phase transition showing the physical extent and entanglement properties of the typical longest entangled cluster in the different regimes. 
  }
  \label{fig:ScalingModel} 
\end{figure}

As alluded to already, the transition from the MBL to the thermal phase appears to be driven by a sparse cluster which looks like a resonant backbone of entangled spins which is \emph{just} able to act as a functional bath for the rest of the system on the thermal side of the transition as $L\rightarrow\infty$. We will use the word ``cluster'' to mean a network of fully or partially entangled spins which need not be spatially contiguous. It is useful to distinguish two quantities for a given cluster: $\ell_E$ denotes the spatial extent of the cluster \emph{i.e.} the maximum physical distance between any two spins on the cluster, while $S_E$ denotes the total entanglement in the cluster defined, say, as the entanglement entropy (in bits) for a cut in the middle of the cluster. Since the cluster could be spatially sparse and its constituents only partially entangled, it is possible for $S_E \ll \ell_E/2 $, where $\ell_E/2$ is the infinite temperature thermal entropy for a cluster of size $\ell_E$. Henceforth, we will refer to $\ell_E$ and $S_E$ of the longest cluster in a typical sample, and we assume that we are close enough to the transition that $\ell_E \gg 1$. Then, we posit that (\emph{c.f.} Fig.~\ref{fig:ScalingModel}):
\begin{enumerate}
\item For large enough $L$ on the MBL side, the system looks strongly localized such that $\ell_E \ll L$ and the typical longest cluster does not span the system. Moreover, the cluster is sparse and weakly resonating so $S_E \ll \ell_E$.

\item In the quantum critical regime, $\ell_E \sim L$ so that the typical largest cluster spatially spans the entire system. However, it is still the case that $S_E\ll \ell_E$ so the cluster looks like a \emph{sparse} network of resonances. Entangled spins that lie on the cluster are ineffectual in thermalizing the rest of the system.

\item If we start in the QC regime with $\ell_E \sim L$ and increase $L$ on the MBL side of the transition (red line in Fig~\ref{fig:ScalingModel}), $\ell_E$ initially grows with $L$. As the (subtle) QC-MBL crossover is approached, the growth of the cluster slows such that it no longer spans the system as we pass through the crossover. Throughout the scan, the network of entanglement remains sparse such that $S_E\ll \ell_E$. It is possible that the sparse critical cluster evolves and matches onto rare Griffiths regions\cite{Gopalakrishnan15,ZhangRG} deep in the MBL regime.

\item If we start in the QC regime and increase $L$ on the thermal side of the transition (blue line in Fig~\ref{fig:ScalingModel}), $\ell_E$ grows with $L$ such that the typical largest cluster continues to span the entire system even as $L$ is increased. As the system approaches the QC-thermal crossover, the sparse cluster starts ``filling in'' by thermalizing the remaining localized regions
and $S_E$ grows (Fig.~\ref{fig:CrossoverModels}b). At the crossover, $S_E$ is some finite and sizeable fraction (say half) of the thermal entropy for the cluster. In the thermal phase, the network of resonances becomes a fully functional bath which is able to thermalize the rest of the system such that $\ell_E  = L = 2 S_E$ for large $L$.

\end{enumerate}

Let us connect this picture to the VHA\cite{VHA} and PVP\cite{PVP} RG analyses. Both VHA and PVP start with a distribution of microscopic thermal clusters interspersed between localized spins. As the RG proceeds by integrating out short-distance physics, clusters can combine and recombine and hence grow in size. VHA's RG allows for both thermal \emph{and} localized blocks to grow, and makes the crucial assumption that when two disjoint thermal blocks get entangled, they thermalize all the degrees of freedom between them.
Thus, in the VHA treatment, $S_E = \ell_E/2$ by construction and the system always looks locally thermal or localized. On the other hand, our picture seems closer to PVP's analysis. In PVP's RG, only thermal clusters are allowed to grow by forming resonances with other clusters. Moreover, they do not insist on fully thermalizing the insulating regions spatially interspersed between the resonating clusters, thereby allowing for a sparse microscopically inhomogeneous backbone.

We now discuss how our picture fits with the observed data. In our picture, 
the sparse network of entangled spins in the QC regime is only subthermally entangled.  Thus, the average EE for small subregions will be subthermal in the QC regime (and at the crossovers) consistent with Fig.~\ref{fig:S1}. As mentioned earlier, PVP's data also show that only a small fraction of all the spins are in thermal clusters at the transition and thus also predicts a subthermal EE for a small subregion, in contradiction with Grover's scaling. Nevertheless, PVP's discussion emphasizes agreement with Grover's thermal scaling in several places, leaving the apparent contradiction unaddressed/unresolved. Of course, our picture also predicts a subthermal half-chain EE at the QC-thermal crossover in agreement with all the different RG treatments\cite{VHA, PVP, ZhangRG}, but this is less surprising since Grover's bounds in Ref.~\onlinecite{GroverCP} do not constrain this quantity.

Since the network of resonances in our picture at the QC-thermal crossover looks homogeneous on the macroscale with strong inhomogeneities only at the micro-level (Fig.~\ref{fig:CrossoverModels}b), $\Delta_S^{\rm cuts}$ is asymptotically less than the maximum allowed value $\sim L$. As an illustrative example of such a discontiguous network, assume that, at the crossover, every site in the chain has equal probability of either belonging to a maximally entangled cluster or not. Since only approximately half the sites in any subregion will be part of the cluster, the average subregion EE will be subthermal.
Moreover, $\Delta_S^{\rm cuts}$ in this model of random occupations clearly gives a $\sqrt{L}$ scaling in agreement with the sub-volume law scaling for $\Delta_S^{\rm cuts}$ in Fig.~\ref{fig:Variance}. On the other hand, VHA's RG predicts 1-2 long locally homogeneous thermal and insulating blocks of size $O(L)$ at the crossover (Fig.~\ref{fig:CrossoverModels}a), giving an $O(L)$ scaling for $\Delta_S^{\rm cuts}$, in contradiction with our observations.  Again, this difference might be due to our data being in a pre-asymptotic regime.

Finally, we note that we have depicted the QC-thermal crossover regime as a wide wedge in Fig.~\ref{fig:ScalingModel}. Different samples can go through the crossover at different values of $W$, thereby giving a large sample-to-sample variation in the half-chain EE at the QC-thermal crossover.  The trend with increasing $L$ towards stronger sample-to-sample variations
is consistent with the observed trend towards bimodality in the distributions of the volume-law-coefficient of the EE in Ref~\onlinecite{ClarkBimodal}. {In fact, the trend towards bimodality near the crossover makes a stronger statement since it indicates that samples lying on either side of the crossover have markedly different entanglement structures consistent with the discontinuity we have discussed.} 
We also note that the Harris/CLO bounds\cite{CLO} do not constrain the intra-sample variations of quantities, say across eigenstates.  This allows the width of the finite-size scaling window in individual samples to be much narrower than the width of the scaling window across samples (which is constrained by Harris/CLO).  {We need this narrowness in the scaling window to meaningfully talk of individual samples being on either side of the crossover within a broad sample-to-sample spread.}
This is reminiscent of the scaling of disordered first-order thermodynamic phase transitions in $d >2$ where the width in individual samples ($\sim 1/L^d$) is much narrower than the width across samples ($\sim 1/L^{d/2}$)\cite{Privman1983, FisherBerker, CCFS2}. 

 \begin{figure}
  \includegraphics[width=.9\columnwidth]{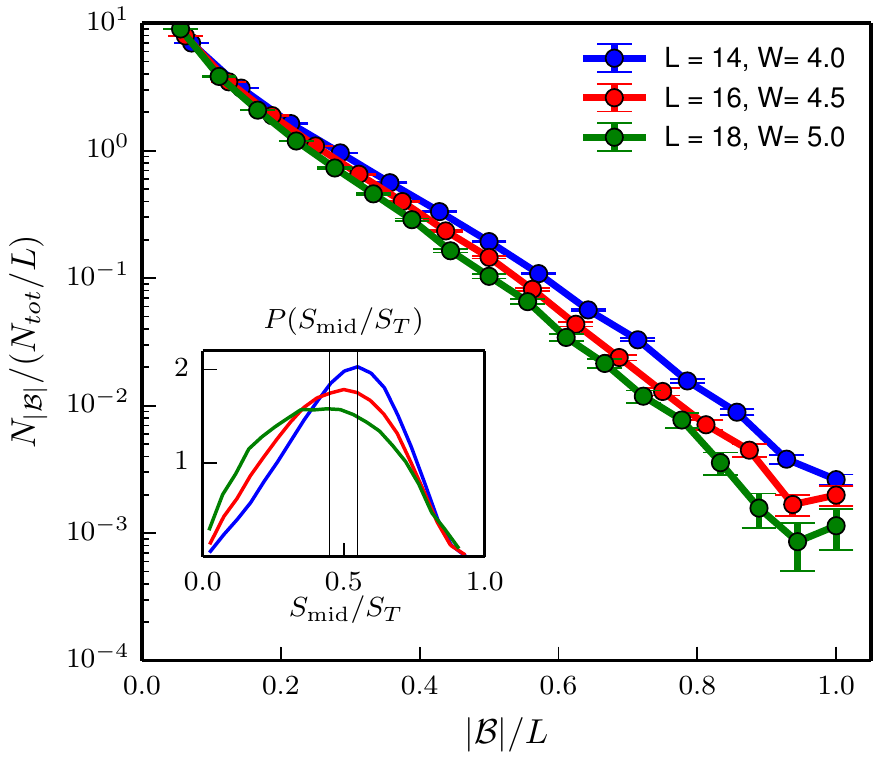}
  \caption{Distribution of thermal block sizes for eigenstates with $0.45<S_{\rm mid}<0.55$. The exponential decrease in $N_{|\mathcal{B}|}$ with $|\mathcal{B}|$ suggests that the local entanglement structure in these states looks inhomogeneous with a network of small interspersed thermal and localized blocks. (inset) Distributions for $S_{\rm mid}$ near the crossover for different $L$s and $W$s.}
  \label{fig:Backbone}  
 \end{figure}

We now present one last piece of numerical data indicating that the entangled clusters in these size ranges are not large contiguous blocks at the crossover. To discriminate between the two proposals in Fig~\ref{fig:CrossoverModels}, we pick values of $W$ near the crossover on the thermal side, $W \sim 4.0-5.0$. The average half-chain EE for a cut in the middle of the system, $S_{\rm mid}$, is roughly $0.5 S_T$ at the $W$ we use for each $L$, although the distribution across eigenstates and samples is fairly broad (Fig.~\ref{fig:Backbone}(inset)). To probe the local spatial structure of entanglement, we pick all eigenstates across all samples with $0.45<S_{\rm mid} /S_T < 0.55$.  We only consider eigenstates in this small range of $S_{\rm mid}$ to avoid the conflating effects of local variations in entropy which are correlated with large/small values of $S_{\rm mid}$.

To obtain the structure of thermal clusters, we compute the single-site $S_1$ for each site in each eigenstate in this restricted set and obtain the median value of $S_1$ denoted $S_1^{\rm med}$. Note that $S_1^{\rm med}$ is chosen once and for all across all states and sites in the ensemble. Then, in each eigenstate, all sites with $S_1 \geq S_1^{\rm med}$, $S_1< S_1^{\rm med} $ are labelled  ``thermal'', ``localized'', respectively.
We define a thermal block $\mathcal{B}$ as a \emph{contiguous} set of ``thermal'' sites so-defined and obtain the lengths $|\mathcal{B}|$ of all blocks. If the picture in Fig~\ref{fig:CrossoverModels}(a) holds, we expect the typical block size to be $O(L)$, whereas the picture in Fig~\ref{fig:CrossoverModels}(b) requires the resonating cluster to be comprised of many small blocks.

Fig~\ref{fig:Backbone} shows a distribution of normalized block sizes where $ N_{|\mathcal{B}|}$ is the number of blocks of a given size $|\mathcal{B}|$ and this quantity has been normalized by the average number of blocks at each size assuming a uniform distribution over sizes.
The data shows an exponential decrease in $N_{|\mathcal{B}|}$ with $|\mathcal{B}|$, supporting a scenario consistent with a sparse, microscopically inhomogeneous entangled cluster.

 \section{Concluding Remarks}
We studied the finite-size quantum critical and crossover regimes of the MBL-to-ETH phase transition and found evidence supporting a view of this transition as a hybrid between continuous and discontinuous phase transitions. We showed that $S_A$, the entanglement entropy of subregions $A$ much smaller than the system size, looks strongly subthermal in the critical regime, contrary to an established constraint which requires $S_A$ to be thermal at the transition if it is continuous\cite{GroverCP}.  This contradiction is resolved by positing that $S_A$ varies \emph{discontinuously} across the transition, thereby violating a crucial assumption in the derivation of the constraint. This is a striking result for a seemingly local property across a transition which otherwise looks continuous in many respects.

We also studied the variance of the half-chain entanglement entropy and parsed in detail the contributions coming from sample-to-sample, eigenstate-to-eigenstate and cut-to-cut variations. Notably, we  observed a volume law scaling for the standard deviation of the half-chain EE across eigenstates of the same sample, while the cut-to-cut variations were found to be sub-dominant. 
We also found that the sample-to-sample variations give the largest contribution and grow strongly (super-linearly) with increasing $L$ at the system sizes studied, a trend that is unsustainable in the asymptotic large-$L$ limit and is consistent with observed violations of Harris-Chayes/CLO exponent inequalities. Our analysis suggests the possibility of two asymptotic fixed points governing the MBL transition: one dominated by ``intrinsic'' intra-sample randomness, and the second dominated by external inter-sample quenched randomness. A deeper of study of this critical structure, say via a comparison to quasiperiodic models with no quenched randomness is an interesting direction for future study\cite{VK_QP}. 

We presented a heuristic picture in which the transition to the thermal phase is driven by an eigenstate-dependent sparse resonant cluster of long-range entanglement, which just barely gains enough strength to thermalize the entire system on the thermal side of the transition as the system size is taken to infinity. This cluster looks strongly inhomogeneous on the microscale, with small interspersed blocks of entangled and localized spins, but has a more homogeneous macrostructure with long range entanglement between separated blocks of spins.  We discussed the evolution of the size and entanglement properties of this resonant cluster across the phase transition, and situated our picture relative to existing renormalization group frameworks for the transition. We explained how discontinuities in local properties like $S_A$, in fact, stem from a global discontinuity --- the ability of the ``backbone'' of entanglement to effectively act as a bath and thermalize the rest of the system.

Going forward, it would be extremely interesting to find a prescriptive way of identifying the dominant entanglement clusters in eigenstates and to  compare their structure with our proposed scenario. It would also be interesting to see the evolution of these clusters across the transition, and whether they connect up with the rare thermal Griffiths regions which dominate the low-frequency dynamics deep in the MBL phase\cite{Gopalakrishnan15, ZhangRG}. Additionally, a more detailed analysis of the finite-size scaling windows, both inter- and intra-sample, is essential for better understanding the properties of this fascinating dynamical quantum phase transition.

\label{sec:conclusion}

{\it Acknowledgements:}
We thank Ehud Altman, Anushya Chandran, Bryan Clark, Trithep Devakul, Chris Laumann, Vadim Oganesyan, Shivaji Sondhi, Romain Vasseur and especially Tarun Grover for stimulating discussions.  This work was supported by NSF grant DMR-1408560 (DS) and by the Addie and Harold Broitman Membership at the I.A.S. (DH).

\begin{appendix}

\section{Distributions of $S_1$}
\label{app:S1dist}

In this appendix, we present data for the distributions of the end spin entanglement entropy $S_1$. These carry more information than the mean values presented in Fig.~\ref{fig:S1}, and provide further evidence in support of subthermal values for $S_1$ in the quantum critical regime. 

\begin{figure}
  \includegraphics[width=\columnwidth]{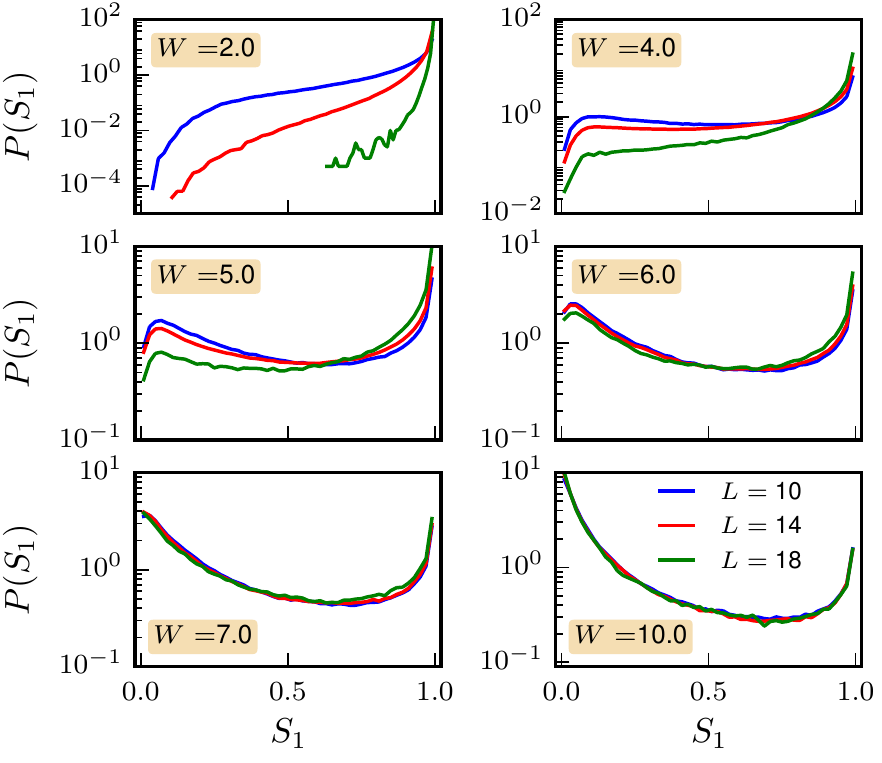}
  \caption{Probability distributions of the end spin entanglement entropy $S_1$ for different $W$s and $L  = 10, 14, 18$ (blue, red, green curves). The distributions become extremely broad near the transition with very little system size dependence, consistent with strongly subthermal mean values of $S_1$ in the quantum critical regime. } 
  \label{fig:S1Dist}  
 \end{figure}

Fig.~\ref{fig:S1Dist} shows distributions of $S_1$ across eigenstates and disorder realizations for different $W$'s and system sizes. 
We find that deep in the thermal phase ($W=2.0$), the distribution of $S_1$ is peaked  near the thermal value of one bit and the distribution becomes significantly sharper with increasing system size (notice the logarithmic scaling on the y-axis). As the transition is approached, the distributions become broader and the system size dependence becomes weaker. Near criticality ($W=6.0, 7.0$) and in the MBL phase, the distributions are extremely broad with virtually no flow with system size. Of course, such broad distributions imply a strongly subthermal mean $S_1$ in accordance with our data in Fig~\ref{fig:S1}.    

\end{appendix}

\bibliography{global}

\end{document}